\documentclass[preprint]{elsarticle}

\usepackage{amsmath, mathrsfs, amssymb,amsfonts,amsthm,graphicx, epsf, dcolumn}
\usepackage[utf8]{inputenc}
\usepackage{amssymb,amsmath,amsfonts}
\usepackage{color}
\usepackage{slashed}
\usepackage{braket}
\usepackage{slashed} 
\usepackage{centernot} 
\usepackage[colorlinks=true]{hyperref}  
\hypersetup{
    bookmarks=true,         
    unicode=false,          
    pdftoolbar=true,        
    pdfmenubar=true,        
    pdffitwindow=false,     
    pdfstartview={FitH},    
    colorlinks=true,       
    linkcolor=magenta, 
    citecolor=blue,        
    filecolor=magenta,      
    urlcolor=cyan           
} 

\journal{Physics Letters B}

\begin{document}

\begin{frontmatter}



\title{A unified geometric description of the Universe: \\
from inflation to late-time acceleration without an inflaton nor a cosmological constant}


\author{Luisa G. Jaime}
\ead{luisa@ciencias.unam.mx}

\author{Gustavo Arciniega}
\ead{gustavo.arciniega@ciencias.unam.mx}


\address{Departamento de F\'{\i}sica, Facultad de Ciencias, Universidad Nacional Aut\'onoma de M\'exico, Apartado Postal 50-542, CDMX, 04510, Coyoac\'an, M\'exico}


\begin{abstract}
We present a cosmological model arising from a gravitational theory with an infinite tower of higher-order curvature invariants that can reproduce the entire evolution of the Universe: from inflation to late-time acceleration, without invoking an inflaton nor a cosmological constant. The theory is Einsteinian-like. The field equations for a Friedmann-Lema\^{i}tre-Robertson-Walker metric are of second-order and can reproduce a late-time evolution that is consistent with the acceleration provided by the cosmological constant at low redshift. Our results force us to reinterpret the nature of dark energy, becoming a mechanism that is inherited solely from the geometry of spacetime.
\end{abstract}




\end{frontmatter}


\section{Introduction}

One of the most elusive and striking fundamental problems in modern physics is the nature of the cosmological constant, $\Lambda$, first introduced by Einstein in 1917 \cite{Einstein1917}, and recovered in 1998 as a fundamental part of the theory by the Supernova Cosmology Project \cite{Perlmutter:1998np} and the Supernova Search Team \cite{Riess:1998cb}. A first attempt to embed $\Lambda$ with a physical meaning of a quantum vacuum expectation value was made by Zel'dovich in 1967 \cite{Zeldovich1967} after the cosmological constant was suggested by the analysis of observations made by Petrosian, Salpeter, and Szekeres \cite{Petrosian1967}. Unfortunately, Zel'dovich's computation started the discrepancy that continues till the present time between the theoretically estimated value for $\Lambda$ and the observed one, from a difference of $\sim 10^{8}$ computed by Zel'dovich to $\sim 10^{120}$ estimated from Quantum Electrodynamics (see \cite{Rugh:2000ji} for a review regarding the quantum interpretation of the cosmological constant). Since then, the problem has been addressed from different perspectives, including the string theory's interpretation from a quantum gravity perspective \cite{Witten:2000zk}, without achieving a satisfactory answer about its nature. Nevertheless, despite the increasing amount of problems related to the existence of a cosmological constant \cite{Perivolaropoulos:2021jda}, the $\Lambda$ paradigm appears to be unavoidable. Furthermore, a non-zero positive value of the cosmological constant is the most vital feature standing from the Swampland conjectures for string theory in cosmology \cite{Agrawal:2018own, Obied:2018sgi, Ooguri:2018wrx, Vafa:2005ui}.

Recently, there appear a series of papers pointing out a geometric scenario for an initial accelerated inflationary universe that transits smoothly to a $\Lambda$CDM-like late-accelerated universe \cite{Arciniega:2018fxj, Cisterna:2018tgx, Arciniega:2018tnn, Arciniega:2019oxa, Arciniega:2020pcy, Cano:2020oaa, LuisaSola1}, without invoking a quantum inflaton field, $\phi$\footnote{Some authors keep the inflaton field as a final stage on a stepdown ladder of curvature invariants  \cite{Edelstein:2020lgv, Edelstein:2020nhg} }. This theory, called Geometric Inflation (GI), inspired us to wonder if it would be possible to do the same for the cosmological constant, \textit{i.e.} to adjudge to a geometric mechanism the late-time accelerated expansion. We found that the answer is yes. There is a way to construct robust cosmological models from the Geometric Inflation Lagrangian densities that gives a viable universe at all times: from an exponential inflationary epoch to a late-time accelerated universe. The former throughout and only using a solely geometric mechanism without the need of an initial quantum inflationary scalar field, $\phi$, nor a cosmological constant.

\section{Cosmology from curvature invariants}\label{cosmoinvar}

The following action describes the theory of Geometric Inflation considering $\Lambda=0$:

\begin{equation}\label{eq:action}
S=\int \mathrm{d}^4x\frac{\sqrt{-g}}{2\kappa}\left\{ R+\sum_{n=3}^{\infty}\alpha_{(n)} \mathcal{R}_{(n)}\right\},
\end{equation}

\noindent where each $\mathcal{R}_{(n)}$ is a Lagrangian density constructed from curvature invariant scalars of order $n$ of Riemann contractions (see the appendix in \cite{Arciniega:2018tnn}), and $\alpha_{(n)}$ encode the energy scale and the coupling constant of the theory for each $n$.

It is possible to demand to the theory to satisfy the following properties:

\begin{enumerate}
\item The linearized equations around a maximally symmetric spacetime are scalar, ghost, and massive-graviton free \cite{Bueno2016, Bueno2016b}.
\item The theory admits non-hairy, spherically symmetric black holes and Taub-NUT/Bolt solutions \cite{Bueno2016a, Hennigar2016, Bueno2017, Cano2019, Bueno:2017qce}.
\item The equations of motion for a Friedmann-Lema\^{i}tre-Robertson-Walker (FLRW) metric are second order for the scale factor $a(t)$ \cite{Arciniega:2018fxj,Arciniega:2018tnn,Arciniega:2019oxa}.
\end{enumerate}

It also has been shown \cite{Arciniega:2018fxj, Arciniega:2018tnn, Arciniega:2019oxa} that the Geometric Inflation theory produces an early exponential acceleration that posses a graceful exit. The theory exhibits a slow-roll inflationary behaviour \cite{Arciniega:2020pcy}, $\epsilon_1=-\dot{H}/H^2 \ll 1$, within the observational constraints imposed by the Planck satellite \cite{Planck2018}. In a recent paper \cite{LuisaSola1}, it has been shown that the Geometric Inflation theory with a cosmological constant and space curvature $k=0$ can provide enough e-fold numbers between the frontier imposed by the Planck density and the exit of inflation for the Horizon and Monopole puzzles. When the inflationary process is ended, the evolution of the Hubble parameter goes immediately to the one that the standard General Relativity (GR) predicts. In this way, the constraints imposed by the Big Bang Nucleosynthesis (BBN) are fulfilled. Finally, from BBN until the present day, the evolution goes as in GR, where the cosmological constant $\Lambda$ drives the late acceleration.

In the present work we consider an homogeneous and isotropic universe with a null spatial curvature $k=0$, an FLRW ansatz, $ds^2=-dt^2+a(t)^2 \left(\frac{dr^2}{1-k r^2}+r^2 d\Omega^2 \right)$, and that fulfills  properties (1-3). The modified Friedmann equations, considering $\Lambda=0$, are given by:

\begin{align}\label{eq:modifried1}
3F(H)&=\kappa\rho,\\ \label{eq:modifried2}
-\frac{\dot{H}}{H}F'(H)&=\kappa(\rho+P),
\end{align}

\noindent where $\dot{H}\equiv \frac{dH(t)}{dt}$, $H'\equiv \frac{dF(H)}{dH}$, $\rho$ is the density energy, $P$ is the pressure, and the function $F(H)$ can be written as:

\begin{equation}\label{eq:Ffunction}
F(H)= H^2+\sum_{n=3}^{\infty} \alpha_{(n)} H^{2n} \, .
\end{equation}

The critical density, $\rho_\text{c}$, is defined as $\rho_\text{c}=3F(H)/\kappa$. In the case that the relation $P=\omega \rho$ is satisfied, then $\rho \propto a^{-3(1+\omega)}$.

\section{The model}

As mentioned earlier, all models analyzed from the Geometric Inflation scheme have been considering the existence of a cosmological constant $\Lambda$. Hence we will refer to the case $\Lambda=0$ as Geometric Inflation with Late-time Acceleration (GILA) to distinguish it from $\Lambda\neq 0$. 

From the expression for $F(H)$, it is convenient for this work to rewrite the equation (\ref{eq:Ffunction}) by splitting the $\alpha_{(n)}$ coefficients into two coefficients $\tilde{\lambda}(n) L^{2(n-1)}$ and $\tilde{\beta}(m) \L^{2(m-1)}$ in the following form:

\begin{equation}\label{eq:Fdarkenergy}
F(H)=H^2+\sum_{n=3}^{\infty}\tilde{\lambda}(n) L^{2(n-1)}H^{2n}+\sum_{m=3}^{\infty}\tilde{\beta}(m) \L^{2(m-1)}H^{2m},
\end{equation}

\noindent where $\tilde{\lambda}(n)$ and $\tilde{\beta}(m)$ are numerical functions of the natural numbers $n$ and $m$, respectively. $L$ and $\L$ are two coefficient numbers, to be determined, that will play the role of the energy scales of the theory. 

In general, if the series expansion of $F(H)$ is written in the following way:

\begin{equation}\label{eq:FgeneralSeries}
F(H)=H^2+\sum_{n=0}^\infty \frac{\lambda^{n+1}}{n! L^2}(L H)^{2(p+nq)}+\sum_{m=0}^\infty (-1)^{m+1}\frac{\beta^{m+1}}{m!\L^2}(\L H)^{2(p+ms)},
\end{equation}

\noindent where, $p$, $q$, and $s$ are fixed integers, and satisfy the condition $q=ks$, with $k$ an arbitrary integer, then, the series converge to the following expression:

\begin{equation}\label{eq:FgeneralDE}
F(H)=H^2+H^{2p}\Big[\lambda L^{2p-2} e^{\lambda (LH)^{2q}}-\beta \L^{2p-2} e^{-\beta ( \footnotesize{\L} H)^{2s}}\Big].
\end{equation}

In equations (\ref{eq:FgeneralSeries}) and (\ref{eq:FgeneralDE}), the constants $\lambda$ and $\beta$ will play the role of coupling gravitational parameters to be determined.


Let us take for example the case $p=4$, $q=2$, and $s=1$ in equation (\ref{eq:FgeneralDE}). From equation (\ref{eq:Fdarkenergy}), we get the following conditions for $\tilde{\lambda}(n)$ and $\tilde{\beta}(m)$:

\begin{eqnarray}\label{eq:condicionconstants}\nonumber
& \tilde{\lambda}\left(2n-3\right)=0, & \quad \tilde{\lambda}(4)=\lambda, \qquad \tilde{\lambda}\left(n\right)= \frac{\lambda^{\frac{n-4}{2}+1}}{[(n-4)/2]!} ,
\\
& \tilde{\beta}(3)=0, & \tilde{\beta}(n)=\frac{(-1)^{n-1}\beta^{n-3}}{(n-4)!}.
\end{eqnarray}


The selected values in the equation (\ref{eq:condicionconstants}) gives the following series when they are substituted in equation (\ref{eq:Fdarkenergy}):

\begin{eqnarray}\label{eq:FDE-model-serie}\nonumber
F(H)&=& H^2+(\lambda L^6-\beta \L^6)H^8+\beta^2  \L^8H^{10}+\left(\lambda^2 L^{10}- \frac{1}{2!}\beta^3 \L^{10}\right)H^{12}
\\  \nonumber 
&& +\frac{1}{3!}\beta^4  \L^{12}H^{14}+ \left(\frac{1}{2!}\lambda^3 L^{14}- \frac{1}{4!}\beta^5 \L^{14}\right)H^{16}+\cdots\, ,
\end{eqnarray}

\noindent which converges to

\begin{equation}\label{eq:FDE-model-1}
F(H)=H^2+H^8\left(\lambda L^6 e^{\lambda \left(L H\right)^4}-\beta \L^6 e^{-\beta\left( \footnotesize{\L} H\right)^2} \right).
\end{equation}

The expression for $F(H)'$ is given by

\begin{equation}
\label{eq:Fprima}
F'(H)=2H\Bigg\{1+H^6\Big[2\lambda L^6\Big(\lambda(L H)^4+2\Big)e^{\lambda(L H)^4} +  \beta  \L^6\left( \beta ( \L H)^2-4\right)e^{- \beta ( \footnotesize{\L} H)^2}\Big]\Bigg\}.
\end{equation}

We explore the model in equation (\ref{eq:FDE-model-1}) because it reduces to the archetypical model in standard Geometric Inflation \cite{Arciniega:2018tnn, Arciniega:2019oxa, Arciniega:2020pcy, LuisaSola1}, when $\beta=0$ in equation (\ref{eq:FDE-model-1}). It is also important to mention that equation (\ref{eq:FgeneralDE}) obeys the same qualitative behaviour regardless of the chosen values for $p$, $s$, and $q$. This choice does not imply a fine-tuning in the selection.

\section{Geometric Inflation and Late-time Acceleration}

In the equation (\ref{eq:FDE-model-1}) the inflationary period occurs when the high energy regime dominates ($H(z) \gg H_0$), while the late-time acceleration takes place at the present day, i.e. at values of the Hubble parameter close to the present value $H_0$ ($ H(z) \sim 1/H_0$).

\begin{figure}[h!]
\begin{center}
\includegraphics[width=8.6cm]{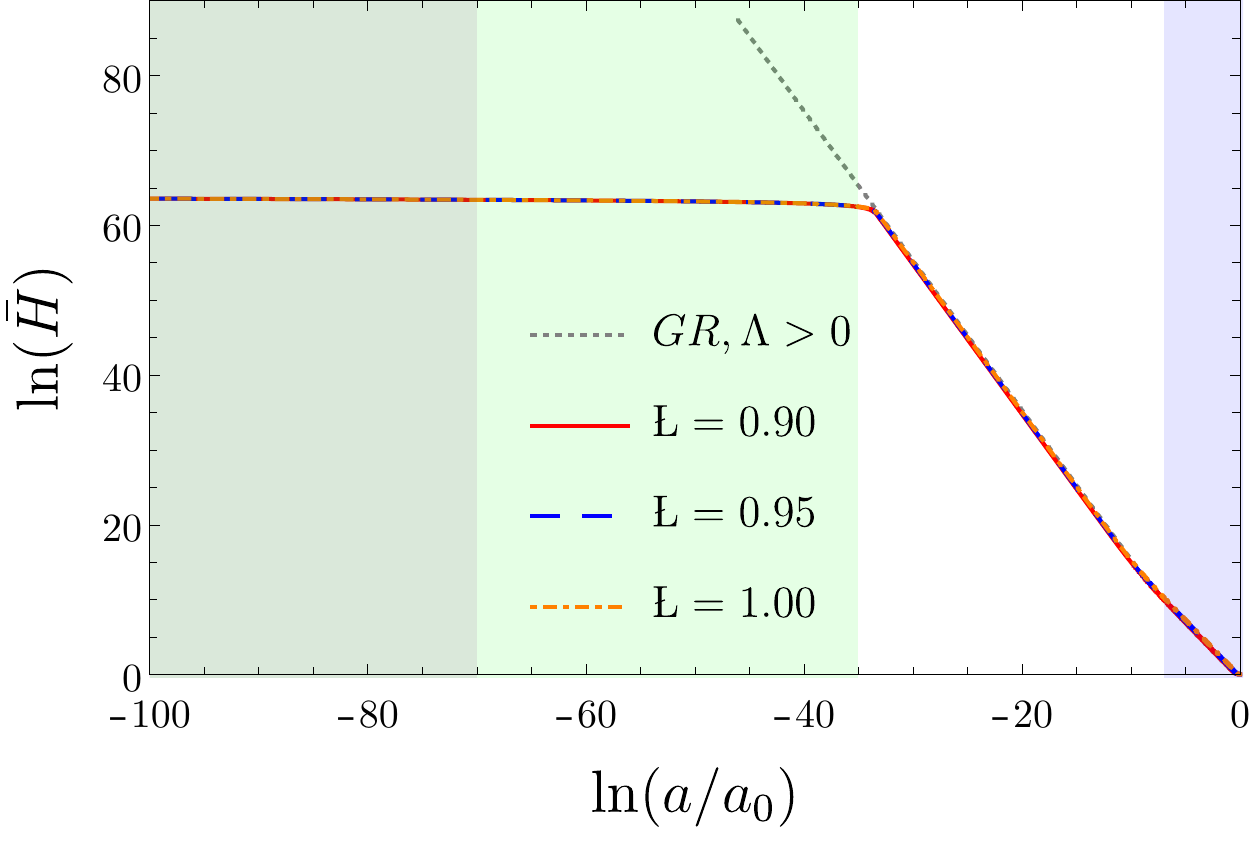}
\caption{Evolution $\bar{H}(z)=H/H_0$. Gray dashed line corresponds to the evolution of $H(z)$ for the General Relativity case with $\Omega_m=0.31$, $\Omega_r=8.4\times 10^{-5}$ and $\Omega_\Lambda = 0.69$. The solid red line, the blue dashed line and the orange dot-dashed line depict the evolution for three cases of the low energy scale, $\L=0.90, \, 0.95,$ and $1.00$, respectively. The high energy scale is fixed to $L=1\times 10^{-27}$. We take the values $\lambda=\beta=1$. In all cases $\Omega_\Lambda=0$ while $\Omega_m=0.999916$, and $\Omega_r=8.4\times 10^{-5}$. The darkgreen area depicts the forbidden zone imposed by the Planck density. The lightgreen area corresponds to the inflationary epoch where $-\dot{H}/H^2 \ll 1$, and the light blue zone start at the size of the Universe when the CMB occurs.}
\label{Fig:HLGA-full}
\end{center}
\end{figure}

In order to explore the cosmic evolution of the field equations (\ref{eq:modifried1}) and (\ref{eq:modifried2}) we solve numerically the differential equation for $\dot{H}$ (\ref{eq:Fprima}). We consider only matter and radiation as the main components of the Universe. The value of the density parameters are: $\Omega_\Lambda=0$ while $\Omega_m=0.999916$, and $\Omega_r=8.4\times 10^{-5}$. We take a fixed value of $L=1\times 10^{-27}H_0^{-1}$ to match with the analysis given in \cite{LuisaSola1}, and three values of the low energy scale $ \L=1.00 \, H_0^ {-1},$ $0.95 \, H_0^ {-1},$ and $0.90 \, H_0^ {-1}$.  Interestingly enough is the fact that the initial conditions can be fixed at the present day, and the integration performed directed to the past, just like in the case of standard GR. Moreover, the fixing of the initial conditions is free of any fine-tuning. In figure \ref{Fig:HLGA-full} we can see the evolution of $H(z)$ during the whole history of the Universe. We notice that the inflationary period takes place until approximately ln$(a/a_0)\sim -35$, when $a(t)$ goes out from the exponential expansion to a graceful connection with the GR regime within some mild differences. Remarkably, the evolution around the present-day exhibits a change in the slope of $H(z)$, producing accelerated evolution at late times, similar to GR with $\Lambda$.

\begin{figure}
\begin{center}
\includegraphics[width=8.6cm]{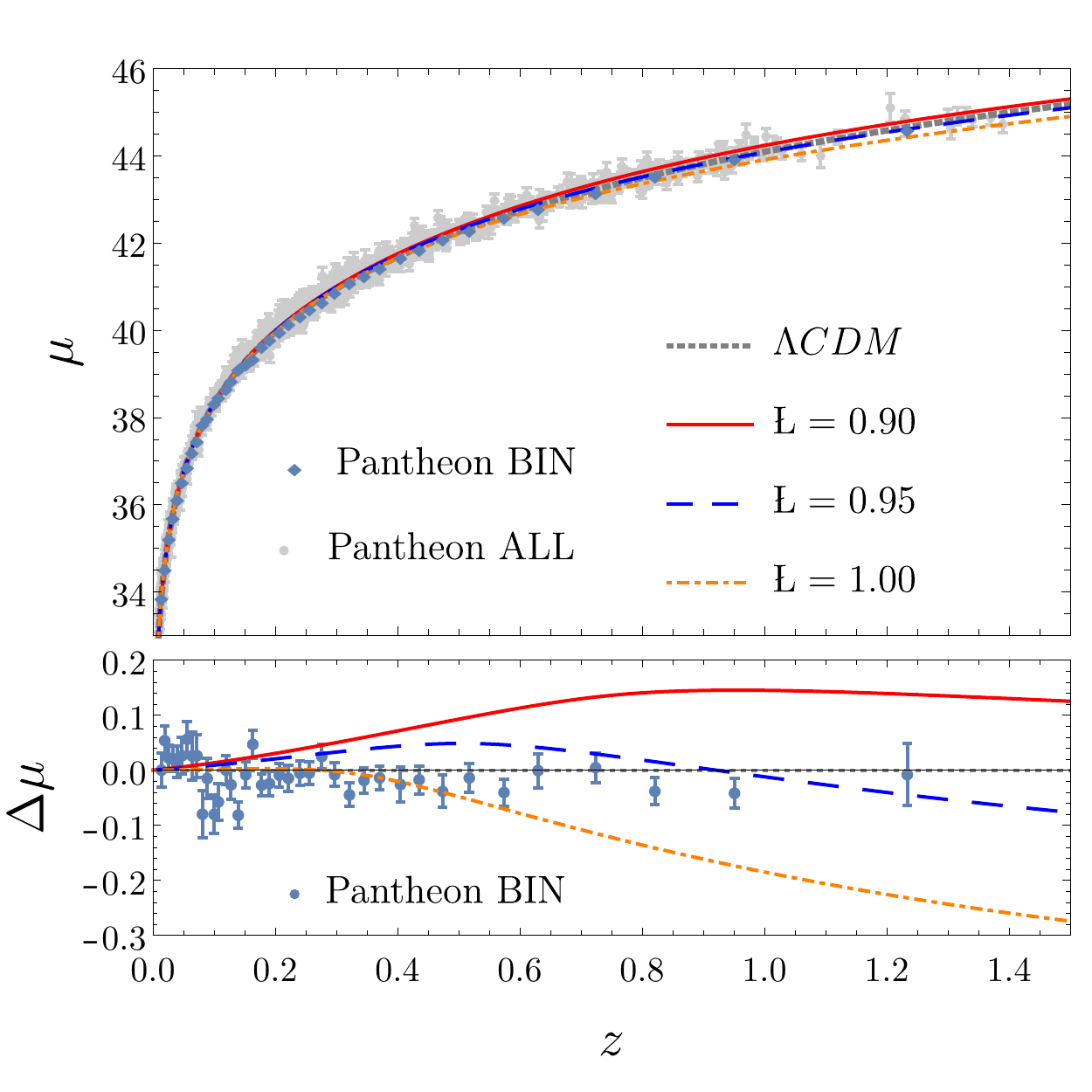}
\caption{Top panel: Plot of the magnitude $\mu$ for the Pantheon sample against redshift, $z$. The dashed gray line shows the standard evolution of GR with $\Omega_m=0.31$, $\Omega_r=8.4\times 10^{-5}$ and $\Omega_\Lambda = 0.69$. The solid red line, the blue dashed line and the orange dot-dashed line depicts the evolution for three cases of the low energy scale, $\L=0.90, \, 0.95,$ and $1.00$, respectively. Bottom panel: Shows the difference of $\mu$ when computed within the frame of the GILA theory compared with the value of $\mu$ predicted in GR.}
\label{Fig:SNe-LGA}
\end{center}
\end{figure}

The latter makes us wonder if this accelerated behaviour can successfully mimic the evolution driven by the cosmological constant.  Figure \ref{Fig:SNe-LGA} shows the  SuperNovae Type Ia Pantheon sample, \cite{Pan-STARRS1:2017jku} where we have plotted the evolution for three different values of $ \L=1,\, 0.95$ and $0.9$. It can be noticed that there is a qualitative agreement.  The bottom panel of Figure \ref{Fig:SNe-LGA} shows the difference of the supernovae magnitude $\mu$ for the GILA theory compared with those in GR,  the binned Pantheon sample takes $H_0=70$ and as a fiducial model, $\Lambda$CDM in GR. We can notice that the low energy term, $ \L$, produces late acceleration going up and down from the acceleration given by the cosmological constant.

\begin{figure}[h!]
\begin{center}
\includegraphics[width=8.6cm]{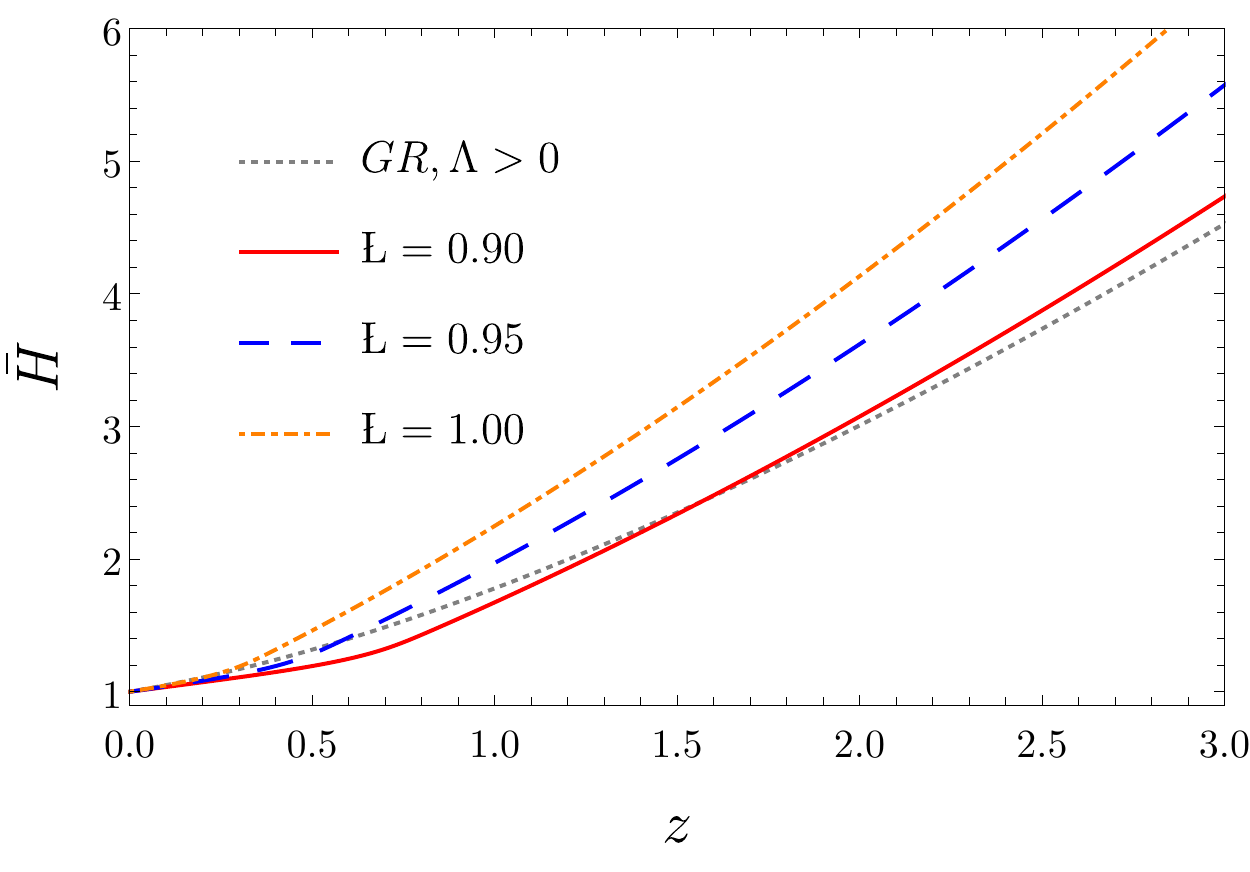}
\caption{Evolution of the Hubble parameter $\bar{H}(z)=H/H_0$. The dashed gray line shows the standard evolution of GR with $\Omega_m=0.31$, $\Omega_r=8.4\times 10^{-5}$ and $\Omega_\Lambda = 0.69$. The solid red line, the blue dashed line and the orange dot-dashed line depicts the evolution for three cases of the low energy scale, $\L=0.90, \, 0.95,$ and $1.00$, respectively.}
\label{Fig:Hlate}
\end{center}
\end{figure}

Figure \ref{Fig:Hlate} explores the evolution of the Hubble parameter at late-time. We can notice that the evolution of $H(z)$ in GILA, for the cases $ \L=0.95$ and $ \L=0.9$ goes below the values of $H(z)$ predicted in GR for low values of $z$ and, as $z$ increases the evolution of the Hubble parameter reaches higher values than those predicted in  GR. 

\begin{figure}[h!]
\begin{center}
\includegraphics[width=8.6cm]{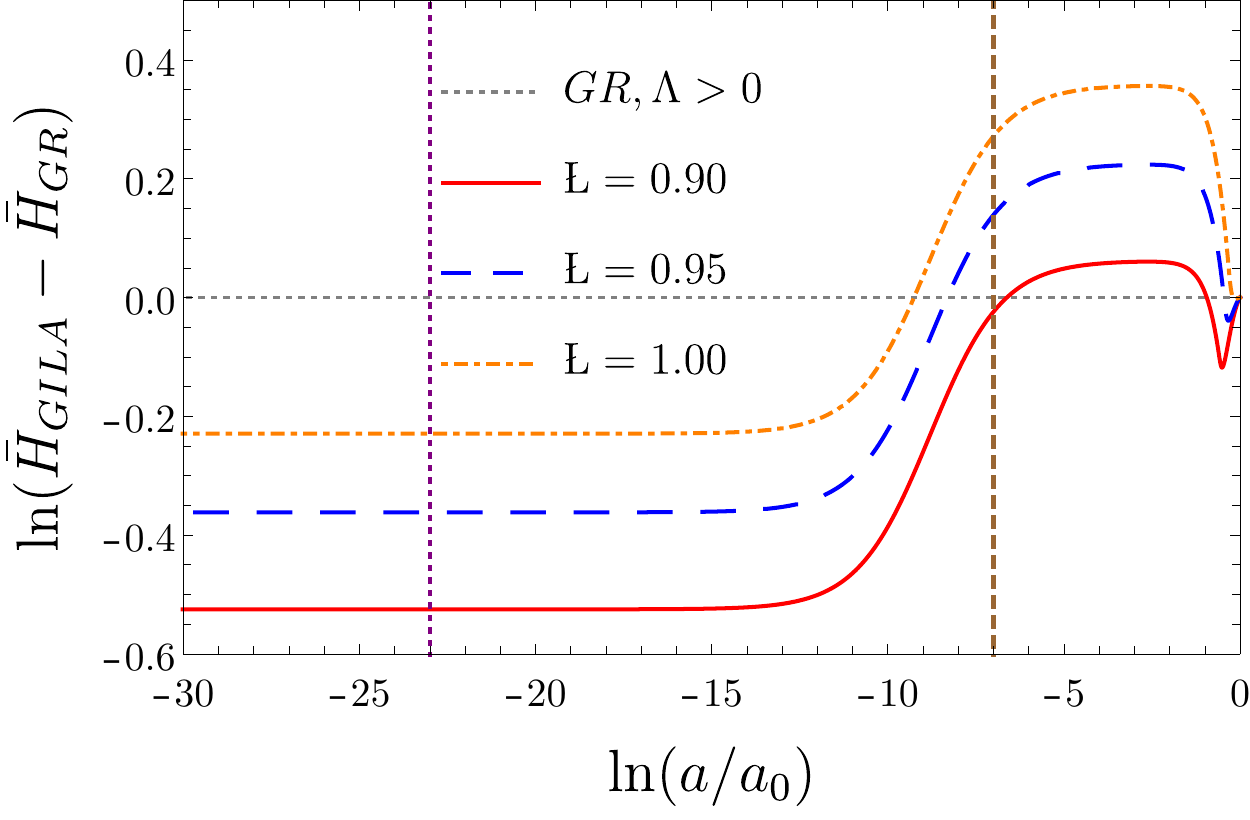}
\caption{Evolution the difference of $\bar{H}(z)=H/H_0$ computed under GILA and the one obtained in by GR, $\Delta(\bar{H}_{GILA}-\bar{H}_{GR})$. The dashed gray line shows the standard evolution of GR with $\Omega_m=0.31$, $\Omega_r=8.4\times 10^{-5}$ and $\Omega_\Lambda = 0.69$. The solid red line, the blue dashed line and the orange dot-dashed line depicts the evolution for three cases of the low energy scale, $\L=0.90, \, 0.95,$ and $1.00$, respectively. The grey dashed vertical line represents the value of ln$(a/a_0)$ at the redshift $z_*$, and the purple dotted line represents the value of ln$(a/a_0)$ when the BBN occurs. }
\label{Fig:Diff-H}
\end{center}
\end{figure}

One possibility that has been explored in the literature as an alternative to relax the $H_0$ tension are the early dark energy models (EDE) \cite{Agrawal:2019lmo, Poulin:2018cxd, Lin:2019qug}. The value of the Hubble parameter around the CMB must be considered to relax the tension. It will have an impact on the size of the sound horizon $r_s(z_*)$ given by

\begin{equation}
\label{eq:rs}
r_s(z)\equiv \int_{z_*}^{z_\text{inf}} \frac{dz}{H(z)\sqrt{3R(z)+1}} ,
\end{equation}

\noindent where $z_*$ is the redshift to the last scattering surface,  $z_\text{inf}$ is the redshift at the end of inflation \footnote{In GR, $z_\text{inf}=\infty $. However, in our theory, it is important to remark that $z_\text{inf}$ is the redshift when inflation ends.}, $z\sim 1089.95$ \cite{Planck2018}, and $R(z)=\frac{3\Omega_r(z)}{4\Omega_b(z)}$.  EDE models provides a lower value of $r_s(z_*)$ by increasing the value of $H(z)$. In figure \ref{Fig:Diff-H} we depict the logarithm of $H(z)_{GILA} - H(z)_{GR}$, where $H(z)_{GILA}$ represents the Hubble parameter predicted by the GILA model. The grey dashed vertical line represents the value of ln$(a/a_0)$ at the redshift $z_*$, and the purple dotted line represents the expected value of ln$(a/a_0)$ at the redshift when the BBN occurs. Depending on the value of $ \L$, the evolution of $H(z_*)$ can take higher or lower values than those obtained in GR. With this, the predicted value of $r_*$ could reach higher values than those obtained in GR. These higher values of $H(z_*)$ provide a way that could help to relax the $H_0$ tension.  It is important to mention that this behaviour does not imply by itself that the $H_0$ tension will be solved. Nevertheless, it is a feature that is naturally provided within this frame, and it is a possibility that is taken by the community in this regard \cite{Knox:2019rjx, Hill:2021yec, Zumalacarregui:2020cjh}.

Regarding the modifications at the BBN time, the value of $H(z)$ is of the same order of magnitude as in GR. Therefore, we do not expect that the primordial abundances of the light elements change significantly from those previously estimated in GR.

\section{Discussion}

We have presented a new gravitational model within the framework of the theory of Geometric Inflation \cite{Arciniega:2018fxj, Arciniega:2018tnn} that can produce inflation, maintaining the features described previously in \cite{Arciniega:2020pcy, LuisaSola1}. The present proposal can produce a late-time acceleration at low energy scales similarly to the one driven by the cosmological constant $\Lambda$. The Hubble parameter evolution presents differences compared to the one predicted in General Relativity. Such differences present features that could be explored as a possibility in relaxing the $H_0$ tension. 

Making a summary of the properties of the theory that we are presenting and exploring to the present day, we can underlay the following ones:

\begin{itemize}
\item[i) ] Its vacuum spectrum consists solely of a graviton, and the theory is ghost-free.
\item[ii) ] It possesses Schwarzschild-like black hole solutions.
\item[iii) ] Its cosmology is well-posed as an initial value problem.
\item[iv) ] It produces a geometric mechanism triggering an inflationary period in the early Universe.
\item[v) ] The inflationary period possesses a graceful exit.
\item[vi) ] The inflationary period can go from the frontier defined by the Planck density to the graceful exit with enough e-folds number to solve the Horizon and Monopole puzzles. 
\item[vii) ] The inflationary predictions, in the slow-roll approximation, are within the constraints imposed by the Planck satellite.
\item[viii) ] The BBN predictions can be fulfilled.
\item[ix) ] The value of $r_*$ can reach lower values than those predicted in the GR framework.
\item[x) ] It produces late-time acceleration without a cosmological constant.
\end{itemize}

It is worth mentioning that, in general, GILA theory is of the fourth order in the field equations. In particular, for spacetimes that differs from maximally symmetric spacetimes, one expects that the second-order equations of motion break down, and one has to deal with fourth-order differential equations mentioned in \cite{Stelle:1977ry, Stelle:1976gc}, with a possibility to renormalize gravity.

Some authors \cite{Pookkillath:2020iqq, Jimenez:2020gbw} have been shown some concerns about instabilities and divergences that appeared when is taking the theory with a finite number of Lagrangian densities into the action.  In this regard, instabilities and other anomalies in perturbations of the theory should be analyzed considering the full infinite tower of Lagrangian densities and the matter content, as has been the case for string theory \cite{Zwiebach:1985uq}. As suggested in \cite{Hohm:2019jgu} for $\alpha'$ corrections, but also valid for an infinite series of gravitational terms: pathologies may be an artefact of the truncation not present in the complete theory. This conclusion is supported by the robustness of the numerical solutions shown in the present and previous works \cite{Arciniega:2018tnn, LuisaSola1}.

Even more, our results have a significant impact on the string theory scenario. The string's Swampland is defined as the ``consistent, anomaly free quantum effective field theories (EFTs) that cannot be embedded in a UV consistent theory of quantum gravity'' \cite{Vafa:2005ui, vanBeest:2021lhn}. That is becoming the most potent argument to disfavour string theory as the correct theory of quantum gravity \cite{Coriano:2020yso}, in particular in cosmology \cite{Agrawal:2018own, Obied:2018sgi, Ooguri:2018wrx}. On the contrary, if the Universe, as we are proposing in this manuscript, has a null cosmological constant value, \textit{i.e.} $\Lambda=0$, the Swampland conjectures do not apply to the string theory as a low energy description of our Universe, and now it is possible to search for an underlying symmetry to explain why is zero the value of the cosmological constant \cite{Bena:2017uuz}.

In this regard, it is worth highlighting the recent work of Hohm and Zwiebach \cite{Hohm:2019jgu, Hohm2019}, where they classified the higher-derivative corrections relevant to cosmology to all orders in $\alpha'$ for a low-energy effective action of string theory. Moreover, they obtained an infinite power expansion in the Hubble parameter $H$ for the Friedmann equations, which is similar to our modified Friedmann equations (\ref{eq:Ffunction}),  opening the possibility that the GILA theory could be a low-energy limit of string theory. 

As a final comment, in the light of the results of the present work, we consider it essential to make a reinterpretation of the role of dark energy as a geometrical characteristic of the gravitational framework for a late-time acceleration era. The former motivates us to explore the possibilities of the theory into other astrophysical scenarios and other theoretical frameworks.

\section*{Acknowledgements}

The authors acknowledge the partial financial support from PAPIIT-UNAM project IN120620, and SNI (CONACYT) LGJ thanks the partial financial support by the CONACyT project 140630.








\end{document}